\documentclass[a4paper,11pt]{article}

\usepackage[english]{babel}
\usepackage[utf8x]{inputenc}
\usepackage[T1]{fontenc}
\usepackage[sc]{mathpazo}  

\usepackage[a4paper,top=2.5cm,bottom=2cm,left=2.5cm,right=2.5cm,marginparwidth=1.75cm]{geometry}

\usepackage{amsmath}
\usepackage{graphicx}
\usepackage[colorlinks=true, allcolors=blue]{hyperref}
\usepackage{wrapfig}
\usepackage[export]{adjustbox}
\usepackage[square,comma,numbers,sort&compress]{natbib}
\usepackage{xspace}

\usepackage{scrextend}  
\usepackage{longtable}
\usepackage{wrapfig}    

\title{Nuclear Physics and the \\
European Particle Physics Strategy Update 2026\\
\vspace*{1cm}
\Large{NuPECC Working Group}}
\usepackage{authblk}

\author[1]{L.M. Fraile} 
\author[2]{J.J. Gaardhøje}
\author[3]{U. van Kolck}
\author[4]{H. Moutarde}
\author[5]{N. Patronis}
\author[6]{M. T. Pe\~na}  
\author[7]{L. Popescu}
\author[8]{V. Wagner}
\author[9]{E. Widmann\footnote{email: eberhard.widmann@oeaw.ac.at}  }

\affil[1]{Universidad Computense de Madrid, Spain}
\affil[2]{Niels Bohr Institute, Copenhagen, Denmark}
\affil[3]{European Centre for Theoretical Studies in Nuclear Physics and Related Areas (ECT*), Trento, Italy}
\affil[4]{Irfu, CEA, Université Paris-Saclay, France }
\affil[5]{Department of Physics, University of Ioannina, Greece}
\affil[6]{Instituto Superior Técnico, Universidade de Lisboa and LIP, Portugal   }
\affil[7]{Belgian Nuclear Research Centre, SCK CEN, Mol, Belgium  }
\affil[8]{Nuclear Physics Institute, Rez, Czech Republic  }
\affil[9]{Stefan Meyer Institute, Austrian Academy of Sciences, Vienna, Austria  }

\date{June 12, 2025}

\usepackage{xcolor}

\newcounter{comment}

\begin{document}
\maketitle

\begin{abstract}
This document provides input to the update of the European Strategy for Particle Physics in fields that are related to Nuclear Physics as described in the NuPECC Long Range Plan 2024 \url{https://arxiv.org/abs/2503.15575}.
\end{abstract}


\vspace*{4cm}
\begin{figure}[h]
\begin{center}
  \includegraphics[width=5cm]{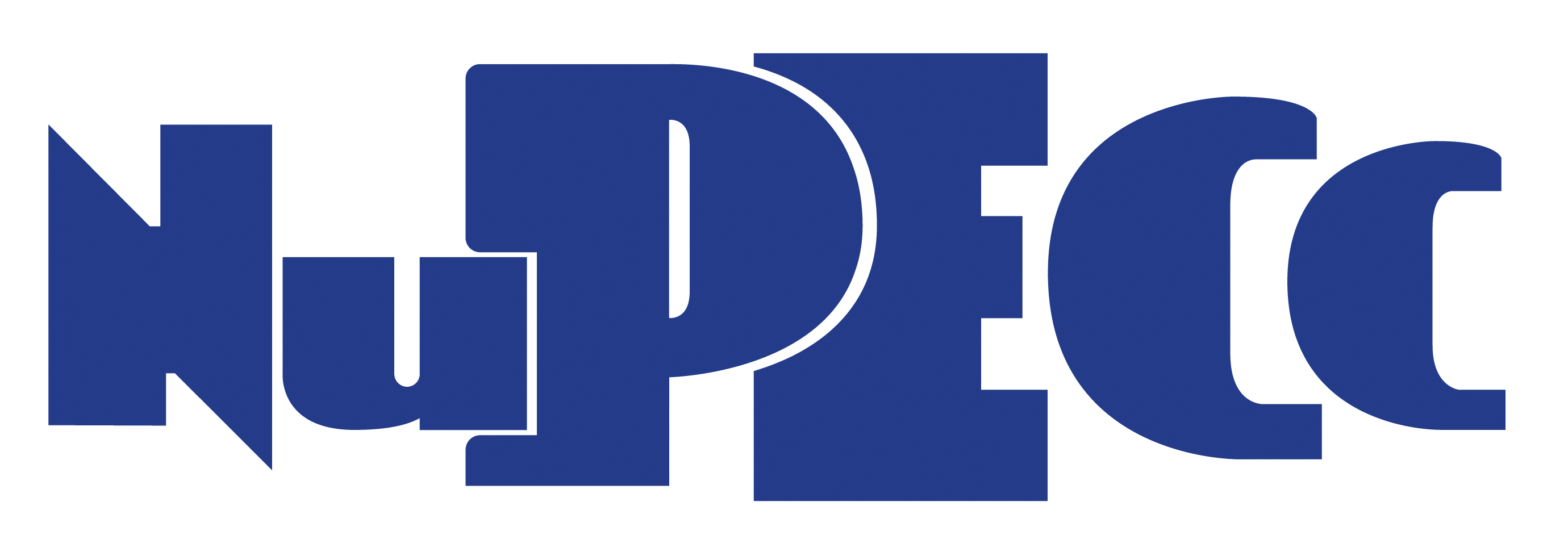}
 
  Nuclear Physics European Collaboration Committee\\
  \url{www.nupecc.org}
\end{center}

\end{figure}

\clearpage\newpage

\section{Introduction}
\label{sec:intro}
Understanding the Universe through fundamental physical principles is a longstanding yet ambitious goal that spans multiple disciplines. Nuclear, particle, and astroparticle physics represented by the NuPECC, ECFA, and APPEC communities, respectively, 
are deeply interconnected fields, each contributing to and complementing the others. Ensuring a rich and diverse physics program beyond the HL-LHC is crucial to address fundamental physics questions, such as the origin of dark matter and neutrino masses, the emergence of visible mass in the universe from the almost massless quarks, the origin and abundance of the elements, the emergence of structure from the non-perturbative strong interaction, to name only a few, in a highly complementary way to colliders. 

Europe is discussing a possible new collider, FCC, Future Circular Collider to succeed the Large Hadron Collider when planned operations stop in 2041 at CERN. If the first phase, an electron-positron collider, is realized this would not permit a heavy-ion program. CERN is encouraged to exploit to the fullest the presently proposed heavy-ion research program including the ALICE-3 detector upgrade. 
CERN should maintain its unique diversity programme by continuing support of non-collider cutting-edge physics projects, both on-site and globally, to promote advancements in particle, astroparticle, and nuclear 
physics. Coordinated efforts, aligned with initiatives such as Joint ECFA-NuPECC-ApPEC symposia and activities, will help maintain Europe’s leadership in these fields. 

Nuclear physics aims to unravel the essential properties of atomic nuclei, from nucleons to quarks and gluons described by Quantum Chromodynamics (QCD). This requires the  understanding of hadronic structure, the residual forces, and the constraints governing the existence of both nuclei and hadrons. Nuclei provide a unique laboratory for studying the strong, weak, and electromagnetic interactions, as well as fundamental symmetries, often complementing particle and astroparticle physics. Additionally, nuclear physics plays a vital role in our understanding of isotope production in astrophysical contexts and in applications across various fields, with strong contributions to the 17 goals of sustainable development adopted in 2015 by the United Nations.


Nuclear physics relies on a multisite approach that requires a variety of state-of-the-art facilities worldwide to advance modern nuclear theory and to produce increasingly unstable radioisotopes and employ advanced technologies and experimental methods. The European NuPECC community profits from a diverse network of infrastructures, from large facilities such as CERN and FAIR to medium-size and small laboratories, which provide the essential particle beams, electromagnetic probes, or neutron sources for research in nuclear physics. 
Specifically important in this context are the proton drivers at CERN that enable the operation of several world-class nuclear physics facilities: AD/ELENA, ISOLDE, n\_TOF, the SPS fixed target program as well as the LHC, 
where a very rich high-energy heavy-ion collision program is located. 
The diversity of infrastructures and the strong theoretical advancements enable research that ranges from fundamental symmetry tests and nuclear structure studies to energy applications and even to medical radioisotope production.


In the following we summarize the physics opportunities and facilities relevant for the European Strategy for Particle Physics Update (ESPPU) from the recently published NuPECC Long Range Plan 2024 (LRP). 
Nuclear physics in this document is understood in a broader sense, covering the topics of Hadron Physics (sec.~\ref{sec:hadron}), Strongly Interacting Matter (sec.~\ref{sec:QGP}), Nuclear Structure and Reaction Dynamics (sec.~\ref{sec:structure}), Nuclear Astrophysics (sec.~\ref{sec:astro}), Symmetries and Fundamental Interactions (sec.~\ref{sec:fun-sym}), as well as Applications and Societal Benefits (sec.~\ref{sec:applications}). Further topics of relevance for the strategy update that are included in this document are People and Society (sec.~\ref{sec:people}) and Theory and Computing (sec.~\ref{sec:theory}).
\clearpage\newpage

\section{Nuclear Physics priorities from the NuPECC Long Range Plan 2024}

\subsection{Hadron physics }
\label{sec:hadron}


The goal of hadron physics is to understand the rich and complex features of the strong interaction in terms of Quantum Chromodynamics. 
This theory plays a particular role in science since, from a restricted set of principles and parameters, it predicts the existence and properties of thousands of particles and countless experimental phenomena. 
Hadron physics is ubiquitous in modern subatomic physics experiments. 
For example, any LHC event is before all a QCD process involving quarks and gluons. 
Most final states reaching detectors are hadrons, whose identification is instrumental to the understanding of the underlying collisions.

The field is driven by several key questions, which remain to this day unsolved problems of physics. 
How does the bulk of the mass of the visible universe emerge from almost massless quarks and massless gluons? 
How do the properties of hadrons emerge from quarks and gluons? 
Is there gluon saturation when gluons are densely concentrated at high energy levels? 
Can massless gluons form massive exotic matter? How does the interplay between QCD short-range interactions and strong couplings lead to multiquark states, such as tetraquarks and pentaquarks? 
What is the role of strong interactions in stellar objects, and in precision tests of the Standard Model? 
Answering these questions requires a diverse set of experimental and theoretical approaches constituting the four
topics of hadron spectroscopy, hadron structure, hadron interactions, and precision measurements. 
Hadron spectroscopy addresses the key question of understanding the observed hadron spectra of the strong interaction and to search for forms of matter beyond the simplest mesons and baryons. 
Hadron structure aims at constructing a comprehensive picture of the spatial distribution and motion of quarks and gluons inside hadrons.
The study of hadron interactions provides insights into the relevant degrees of freedom at different energy scales, as well as the residual interactions between them.
Precision hadron physics is essential in tests of the Standard Model and searches for physics beyond it.

One single experiment cannot address all questions and experimental hadron physics requires a combination of multi-purpose facilities, and smaller experiments with specific aims. 
At present, European hadron physicists successfully conduct experiments at facilities within and outside Europe. 
The NuPECC LRP recommends continuing support to hadron physics programs in existing dedicated facilities (\textbf{AMBER} at CERN, \textbf{HADES} at GSI, \textbf{ELSA} in Bonn, \textbf{MAMI} and \textbf{MESA} in Mainz, \textbf{Jefferson Lab} in Newport News) which have successfully produced a wealth of data. 
Multi-purpose facilities like \textbf{Belle II}, \textbf{BES III} and the \textbf{LHC} have also been decisive in pushing forward our understanding of the strong interaction; support to the ongoing hadron physics activities in these facilities should be maintained. 
At last, in a 10-year perspective, future facilities like \textbf{FAIR} and \textbf{EIC} open new avenues for ground-breaking discoveries and are recommended by the NuPECC LRP. 
They will explore complementary physics cases and can act as unifying forces for the community. 
Moreover, the support of the participation to European researchers to these world-leading facilities will reinforce scientific and technological activities which synergize with future European particle physics projects.



\subsection{Properties of strongly interacting matter}
\label{sec:QGP}


Ultrarelativistic heavy-ion physics at the energy frontier and the field of high density QCD fall in some countries under the overall heading of ‘particle physics’ and in others under the heading ‘high energy nuclear physics’. In some countries that distinction is not made. The recent Long Range Plan from NuPECC identified the following priorities under the heading of ‘Properties of Strongly Interacting Matter at Extreme Conditions of Temperature and Baryon Number Density‘.

The dedicated heavy ion detector, \textbf{ALICE} at CERN, is the primary instrument for unravelling the properties of the Quark Gluon Plasma (QGP), the matter of the early Universe,  using a multitude of detailed probes. The research establishes new limits on fundamental parameters for the QGP, such as its viscosity, flow and correlations, density, temperature and the interaction of heavy quarks in the medium, by colliding heavy nuclei of lead ($^{208}$Pb). Investigations have progressed to studying the collective properties of the QGP, down to p+p collisions, in order to determine a lower formation limit of QGP, also exploiting intermediate collision systems (eg., Xe+Xe). It is tantalizing that the precision in experimental studies is in part set by the understanding of the initial conditions of the colliding nuclei, i.e., its deformation etc. previously thought to be the realm of nuclear structure studies close to the ground state. The development of various collision systems by the LHC is of high interest in order to map out these properties. The ALICE studies are supplemented by studies focusing mostly on high-momentum probes carried out by the \textbf{ATLAS} and \textbf{CMS} collaborations. 

A proposal for a new \textbf{ALICE-3} detector exploiting new technology based on super-thin, flexible, stitched Si detectors has led to the concept of a Si based detector with very low material budget inserted in a new superconducting toroidal magnet. This a major upgrade project for an extended (6 year) Run 5. 
A smaller upgrade (FoCAL), partly anticipating physics at the \textbf{EIC}, aims at measuring gluon properties at low-x by studying small angle electromagnetic decays from Run4 onwards. The \textbf{LHCb} upgrade II will open up for an extended fixed target program and precision studies of the decay of charm and beauty hadrons in heavy-ion collisions.

The phase diagram for strongly interacting matter is rich and essentially still unexplored. Moving to lower energies, away from vanishing chemical potential (particle and anti-particles produced in near equal quantities) the Compressed Baryonic Matter (\textbf{CBM}) detector at FAIR, GSI, will explore the baryon rich domain (finite baryon chemical potential) and search for the critical point for nuclear matter and a possible first order phase transition between the Hadron Gas and the QGP regimes, when SIS100 will start operations. FAIR experiments will also utilize the \textbf{R3B} spectrometer and the \textbf{HADES} Facility.

At CERN’s SPS facility the proposed \textbf{NA60+} experiment will focus on rare lepton decays utilizing its high-rate capabilities to achieve 3-5\% accuracy, and the \textbf{NA61/SHINE} experiment has upgraded its DAQ capabilities for investigating hadron production around the expected onset of deconfinement in a variety of collision systems. 


One of the main topics of discussion around the European Particle Physics Strategy-Update is which next flagship collider to recommend for Europe and CERN. Currently, the discussion focuses on the \textbf{FCC-ee}, a high luminosity electron-positron collider, to be followed by \textbf{FCC-hh}, a hadron collider. If Europe goes for FCC-ee as the first phase it will essentially imply the end of ultrarelativistic heavy ion physics at the energy frontier in Europe (and globally) for several decades. Indeed, there would be a gap of several decades from the shutdown of the HiLumi-LHC (year 2041) until an FCC-hh could become operational (around 2065), implying a major realignment of the (presently) substantial community (approx. 2000 users) working at CERN, although there may be interest in part of the heavy-ion community to study high-multiplicity events also in e$^+$e$^-$ collisions and to pursue fixed target experiments at SPS.

\subsection{Nuclear structure and reaction dynamics}
\label{sec:structure}
The main challenges in nuclear structure and reaction dynamics over the next decade relate to addressing fundamental questions that lie at the heart of modern nuclear physics. The critical objective as proposed by NuPECC is the understanding of how nuclei and nuclear matter arise from the underlying fundamental interactions described by QCD. Determining the limits of nuclear existence and exploring how far the nuclear chart can extend, are also of the utmost interest, so is investigating the exotic phenomena that emerge in open quantum systems where particles can escape or be absorbed.
Other areas of strong interest include the evolution of the nuclear shell structure far off stability, where energy levels shift and adjust as neutrons and protons are being added, and the development and coexistence of the shapes that nuclei can assume from spherical and ellipsoidal forms to more exotic configurations such as pear-shaped or hyperdeformed states.

Understanding the role of nuclear correlations, including pairing interactions and collective excitations, is essential for building a comprehensive picture of nuclear dynamics. The advances in understanding nuclear structure and reaction dynamics will not only refine our understanding of the atomic nucleus but also make significant contributions to broader fields such as astrophysics, through insights into stellar nucleosynthesis and compact objects, see subsection \ref{sec:astro}, hadron physics \ref{sec:hadron}, and the study of fundamental symmetries that govern the universe at its most basic level.

On the experimental side, strong support for both large-scale and small-scale facilities is essential, since it guarantees access to a broad and diverse community, but also facilitates developing and testing instrumentation, and exploratory experiments, which are crucial for training new physicists and preparing for more complex investigations. 
The coordinated efforts among ISOL facilities in Europe have been instrumental in maintaining a world-leading position in radioactive beam science, and further reinforcement of this collaboration is strongly recommended by NuPECC. The unique CERN hadron drivers, due to their energy and intensity, have been essential in this respect to position \textbf{ISOLDE} and \textbf{n\_TOF} as world-leading facilities. Maintaining Europe’s leadership in the use of heavy-ion storage rings as precision instruments through the construction of future rings at \textbf{FAIR} and \textbf{HIE-ISOLDE} are of the highest priority. 

Advanced instrumentation continues to be of the highest importance. In particular, the completion of the \textbf{AGATA-4$\pi$} gamma spectrometer is required for advancing spectroscopy and lifetime measurements, profiting from exceptional resolution and high efficiency. 
AGATA is expected to remain an indispensable tool for nuclear structure gamma-spectroscopy and for precision studies in nuclear astrophysics at both radioactive and stable ion-beam facilities.


NuPECC urges for the completion of the key European facilities including \textbf{FAIR}, \textbf{SPIRAL2}, \textbf{SPES}, \textbf{ELI-NP}, \textbf{ISOL@MYRRHA}, and the planned upgrades at ISOLDE. These laboratories will provide unique insights into the reactions of very exotic nuclei and extend the exploration of the nuclear chart, thereby serving as indispensable platforms for advancing our understanding of nuclear structure and reaction dynamics. 


\subsection{Nuclear astrophysics}
\label{sec:astro}
Nuclear astrophysics is the investigation of nuclear processes that take place in diverse astrophysical environments, exploring their fundamental causes and their impact. The field encompasses a broad spectrum of physical phenomena, from the fusion reactions powering stars to the complex processes driving stellar explosions and cosmic evolution. It  provides the underlying scientific link between different observations of astrophysical objects, offering a unified framework to interpret and understand the universe. A paramount example is our Sun, where recent advancements in understanding the nuclear processes occurring from its core to its outer atmosphere have significantly enhanced our knowledge, not only of solar physics but also of broader areas such as Big Bang nucleosynthesis, cosmic element formation, and space weather.

Furthermore, nuclear astrophysics helps to understand the synthesis of  elements and the chemical evolution of the universe, making it possible to investigate the evolution of elemental abundances throughout the Universe history, and of the properties of dense nuclear matter both in the laboratory and in astrophysical scenarios. In particular, it enables the study of transient phenomena. The advent of gravitational wave telescopes has opened a new frontier in astrophysics, allowing the study of phenomena such as neutron stars and black hole mergers, \textit{kilonovae}, supernova explosions, and gamma-ray bursts using multiple channels. These multi-messenger studies require a solid foundation in nuclear physics, both from the point of view of measurement of nuclear properties and from the theoretical side. In this context the next generation of gravitational wave telescopes (\textbf{Einstein Telescope}) and space-based observatories (\textbf{LISA}) will greatly enhance research in nuclear astrophysics, together with future missions such as \textbf{COSI}, \textbf{eXTP}, and \textbf{e-ASTROGAM}. 

The experimental measurement of processes that are relevant in nucleosynthesis necessitates access to cutting-edge radioactive-beam facilities. The Super-FRS at \textbf{FAIR}, the \textbf{ISOLDE} (CERN) facility upgrades, and \textbf{SPIRAL2} will play an essential role for studying exotic nuclei in explosive stellar events, and their construction and upgrades have been defined as a priority by NuPECC. 
The \textbf{n\_TOF} (CERN) facility will continue to play a crucial role in the study of s-process nucleosynthesis, which is responsible for producing approximately half of the chemical elements heavier than iron. The upcoming beam and detector upgrades will further enhance the facility’s research capabilities.

Small-scale facilities are crucial for nuclear astrophysics research, especially for direct measurements down to the lowest possible energies, as well as indirect measurements, which are strongly supported by NuPECC. A new accelerator mass spectrometry facility with 10 MV or more is required to provide high abundance sensitivity for demanding astrophysical applications.

Additionally, support for the European underground laboratories (such as LNGS \textbf{Bellotti Ion Beam Facility} and \textbf{Felsenkeller}, including its planned DZA upgrade) is essential. 
Dedicated facilities at large laboratories should be fully exploited, such as the \textbf{CRYRING} and \textbf{ESR} storage rings at \textbf{FAIR}. 
Future directions should also include strengthening nuclear astrophysics networks in Europe such as ChETEC-INFRA, and ensuring their sustainability to connect with international networks. The evaluation and accessibility of nuclear and nuclear reaction data with quantified uncertainties are crucial. 


\subsection{Symmetries and fundamental interactions}
\label{sec:fun-sym}


Nuclear Physics has played a major role in finding and establishing the laws which govern Nature at the most fundamental level. 
Thus Nuclear Physics contributes to the studies of symmetries and fundamental interactions, which are key topics in particle physics, in a complementary way to experiments at higher energies.
Among the most notable examples are the discoveries of parity (P) and charge-parity (CP) violations, which triggered intense research on symmetry violations in general.
 
In the next decade, various nuclear physics experiments, including searches for electric dipole moments (EDMs), precision studies of radioactive molecules and beta decay, will push CP violation testing to the next level. Simultaneously, the accuracy of tests for Lorentz/CPT will be boosted by more precise studies of matter-antimatter systems, such as comparing hydrogen and antihydrogen or studying the properties of neutrinos and antineutrinos. These precision tests explore facets of the four fundamental interactions that are not always accessible through colliders. They are conducted in smaller-scale facilities, including laboratories, reactors, and accelerators producing and manipulating highly charged ions, neutrons, muons, antiprotons, and radioactive ion beams. These experiments provide unique access to precision measurements of fundamental constants, such as masses or radii of particles. Teams of physicists and engineers compete with ingenuity to develop a varied range of techniques and probes, which drive this research forward. Quantum Electrodynamics (QED) is tested to high precision in highly charged ions and exotic atoms. The fundamental nature of the weak interaction is probed by scrutinising the beta decay of neutrons and radioactive ions. 
Nuclear physics has played a major role in these advancements and it will continue to do so, offering new, unique, and complementary insights into all known interactions -- electromagnetic, weak and strong -- and also into the gravitational behaviour of neutrons and antimatter.


Despite the success of the Standard Model of Elementary Particle Physics, various observations, such as, e.g., the likely existence of Dark Matter and the Baryon Asymmetry of the Universe, point to the need for its extensions: these call for baryon number  violation, connected or not to lepton number  violation, and for additional CP violation. Symmetry tests broadly can be arranged into (i) precision determination of SM parameters and (ii) searches for physics beyond the SM. They involve the full spectrum of particles from neutrinos and charged leptons to hadrons and stable as well as radioactive ions and molecules. 

Some of the most important facilities in this field are operated by CERN and require the continued availability of high-intensity proton beams. At the  \textbf{AD/ELENA} facility producing worldwide unique beams of low-energy antiprotons, experiments are performing spectroscopy of antihydrogen and antiprotonic helium (ALPHA, ASACUSA), tests of antimatter gravity (ALPHA, AEgIS, GBAR), comparison of proton and antiproton properties (BASE), and the creation of antiprotonic unstable nuclei to study nuclear structure effects (PUMA). Notably first results of the spectroscopy and gravitational interaction of antihydrogen have been achieved and promise most stringed tests of matter-antimatter symmetry in the future. The radioactive beam facility \textbf{ISOLDE} is one of the leading facilities in the world producing unstable nuclei by the ISOL technique for a variety of nuclear physics studies.
Other essential existing facilities in Europe are located at the ESFRI facilites \textbf{GSI/FAIR} (both for unstable nuclei and highly charged ions) and \textbf{GANIL-SPIRAL2}, at other radioactive beam facilities as well as cold and ultra-cold neutron facilities and PSI. 


Future European flagship facilities and experiments that should be supported are 
the radioactive beam facilities \textbf{ISOL@MYRRHA} 
a fundamental neutron physics beam line at \textbf{ESS},
and the future \textbf{CR} and \textbf{HESR} storage rings at \textbf{FAIR} providing unique opportunities by extending the storage ring programmes with highly charged ions to high energies.
Further, a dedicated neutron beamline at the ESS for fundamental particle and nuclear physics will enable experiments exploring the high-intensity frontier, opening opportunities for world-leading precision studies as well as searches for baryon number violation and other beyond standard model phenomena.


\subsection{Applications and social benefits}
\label{sec:applications}

The United Nations adopted in 2015 the 2030 Agenda for Sustainable Development, which led to 17 Sustainable Development Goals (SDGs). 
Those are meant as a call to action for all the governments across the globe, but also for all research communities to contribute. 
Nuclear science and its applications can play a major role in the domains of energy, health, and space. In addition to SDG7 (energy), clean and affordable nuclear energy that is available anywhere contributes to SDG1 (no poverty), SDG8 (economy), SDG9 (industry), and SDG10 (reduced inequalities). Imaging and therapies arising from nuclear physics are commonly used in clinical practice around the world, contributing to SDG3 (health). Nuclear physics techniques such as isotopic markers to study plants and the water cycle have strong effects on SDG2 (zero hunger), SDG6 (clean water), SDG13 (climate action), SDG14 (life below water), SDG15 (life on land). Nuclear physics applications require a highly educated and inclusive workforce, contributing to SDG4 (education) and SDG5 (gender equality). The responsible treatment of nuclear waste from medical and energy applications addresses SDG11 (sustainable cities) and SDG12 (responsible consumption). The nuclear physics-based monitoring of non-proliferation aims to address SDG16 (peace). Finally, the strong collaborative nature of nuclear physics in particular in Europe supports SDG17 (partnership).

The NuPECC LRP recommends to keep on working on improving nuclear data, including both the measurement and the evaluation of nuclear data. Those are needed to support research in the fields of energy, health, space, and material science. 
The interdisciplinary nature of these applications requires
capacity building in the fields of radiochemistry and radiobiology, as well as maintaining nuclear science competencies in Europe. 
The large array of nuclear application research fields will benefit from adapted beam access models reflecting the dynamic developments in these fields.
New generations of nuclear energy sources and the management of nuclear waste through partition and transmutation, depend on sustained technological developments in the present facilities, as well as the completion of \textbf{MYRRHA} and \textbf{IFMIF-DONES}. 
The potential of novel medical radionuclides can only be realized by upscaling the production capacity in Europe to clinically-relevant activities. 
This should be achieved through the enhanced use of the \textbf{MEDICIS} separator at CERN and other \textbf{PRISMAP} emerging facilities.
Europe’s capacity in space dosimetry, radiobiology, and radiation hardness testing requires the sustained effort of the present irradiation facilities, as well as the completion of the first galactic cosmic ray simulator in Europe at \textbf{GSI/FAIR}.
There is an increased need for isotope-sensitive techniques in environmental, heritage, and material science.

\subsection{People and society }
\label{sec:people}

The NuPECC LRP recommends that European funding bodies and institutions see nuclear science as a critical societal investment to inspire the public about nuclear science, train the next generation of nuclear scientists, and support equitable career progression with an inclusive approach to diversity across academic, industrial, and vocational career paths.
The promotion of the education of the young generations, and the support of young researchers in all stages of their scientific career, is an asset for the benefit of the European societies. 
The scientific community must invite the present and future generations into career options and development opportunities in the nuclear sciences and across society, supporting their integration into a diverse and inclusive work environment.

The community of European nuclear physicists, in collaboration with funding bodies and other stakeholders, should support the training of new generations of nuclear scientists, to provide the broad skills required across experimental and theoretical nuclear physics research, as well as all disciplines and industries in our society, commonly relying on expertise, techniques and skills from the nuclear sciences. We emphasize that a key rationale for beyond-collider projects
across various facilities and parallel explorations is their long-term impact on careers
and training. This is essential to prevent an otherwise inevitable generational
knowledge gap.

The NuPECC LRP further supports respectful, inclusive and safe work and training environments in academic, industrial, and vocational nuclear-science careers. 
Recognition and visibility should be given to the critical contributions of early career researchers, in order to foster and attract talent.
In addition, the NuPECC LRP recommends that the network of research organisations, funding agencies, as well as scientific collaborations and conference committees should promote a diversity charter, such as the one prepared by NuPECC together with APPEC and ECFA. 
Diversity should be understood as the acknowledgement and respect of the reality that people differ in visible or invisible ways.
Diversity in the nuclear physics community should mirror the European society we live in.


\subsection{Theory and computing}
\label{sec:theory}

Nuclear theory grounds the various subfields of nuclear physics on the Standard Model of particle physics and provides essential input for the search for new physics. Substantial progress has been made in elucidating how the properties of hadrons and nuclei arise from QCD, 
often relying on progress in computing hardware and algorithms. 
To match experimental progress,
sophisticated approaches need to be developed, encompassing theoretical developments, software development and dissemination, progress in algorithms, and access to European computational infrastructures. 

Hadron physics is challenging because it deals with the strong interaction in its non-perturbative regime, which opens up deep questions on the nature of quantum field theories. In response to a wealth of new data, the spectrum, structure, and interactions of hadrons are studied with advanced amplitude analyses combined with lattice QCD simulations, functional methods, and effective field theories (EFTs).
Thanks to the development of powerful renormalization-group methods for the construction of effective interactions, increasingly complex nuclei are becoming accessible to a systematically improvable solution of the many-body problem starting from EFT interactions. Establishing efficient interfaces between different theoretical frameworks is mandatory, including bridging {\it ab initio} and phenomenological methods and improving many-body methods, time-dependent techniques, and reaction calculations. 
Microscopic models for nuclear structure, decay, reactions, and the equation of state of dense matter are necessary for understanding element synthesis and cosmic chemical evolution. 
Data collected in recent and future experiments with ultrarelativistic heavy ions pose qualitatively novel challenges for our understanding of QCD, which requires progress on pre-equilibrium dynamics, QCD thermodynamics and hydrodynamics, hard and electromagnetic processes in the quark-gluon plasma, and collision modeling.

The increasingly unified and coherent description of strong interactions is crucial for precision tests of the Standard Model and its accidental symmetries. Continuous progress in EFTs across energy scales makes possible a meaningful comparison between colliders and low-energy searches in terms of their sensitivity to new physics. Recent progress quantifies a wide range of observed anomalies and possible new signals, including: lattice QCD and dispersion analyses of the anomalous moment of the muon and the nucleon electric dipole moment; EFT and dispersion estimates of radiative corrections in single-beta decay to constrain exotic interactions; {\it ab initio} and phenomenological calculations of neutrino and dark-matter scattering on nuclei, of nuclear matrix elements of neutrinoless double-beta decay, and of nuclear electric dipole and Schiff moments; interface with atomic and molecular physics for a host of precision tests of QED, 
time-reversal, Lorentz/CPT invariance, etc.

The diversity and the quantity of accumulated and anticipated experimental data require a dedicated theoretical effort. Theorists play an essential role not only in interpreting experimental results but also in providing input and predictions for new experiments. Support for experiment and education is enhanced by emerging virtual access facilities, such as those built by the European networks STRONG-2020 and EURO-LABS. 
The evaluation and accessibility of nuclear and nuclear reaction data with quantified uncertainties, such as provided by recently developed (e.g. Bayesian) methods, are crucial. 
The NuPECC LRP recommends the strengthening of networks and virtual access facilities in Europe (such as ChETEC-INFRA for nuclear astrophysics) to ensure their sustainability and connection with international networks. 

Access to large, fast supercomputing facilities, including state-of-the-art quantum computing platforms, is vital for advanced nuclear physics calculations and multidimensional astrophysical simulations. 
The NuPECC LRP encourages the development, maintenance and dissemination of optimized algorithms and codes, as well as collaboration with IT experts. The use of artificial intelligence and machine-learning techniques, such as Bayesian neural networks, is increasing and promises to speed up numerical calculations significantly.

The NuPECC LRP expresses strong support for the \textbf{theory centers}, which house approximately 1200 nuclear theorists throughout Europe and provide career prospects for young researchers. Stronger pan-European support is needed to ensure that the \textbf{European Centre for Theoretical Studies in Nuclear Physics and Related Areas} (ECT*, Trento, Italy) continues to provide a unique venue for interdisciplinary activities --- including those at the interface with particle physics --- which play a strategic role in the development of nuclear physics in the continent. Centers such as ECT* enable the community to benefit from European investments in experimental, supercomputing and quantum computing infrastructures.




\clearpage\newpage
\section*{Glossary}

\begin{longtable}[l]{ c  c }%
    \raggedleft   
    \begin{tabular}{ll}
        AD & antiproton decelerator \\ 
        AGATA & Advanced GAmma Tracking Array \\ 
        ALICE & A Large Ion Collider Experiment \\ 
        AMBER & Apparatus for Meson and Baryon Experimental Research \\ 
        AMS & accelerator mass spectrometry \\ 
        APPEC & Astroparticle Physics European Consortium \\ 
        ASIC & application-specific integrated circuit \\ 
        ATLAS & particle physics experiment at LHC \\ 
        BELLE & experiment at the KEK B-factory \\ 
        BES & Beijing Spectrometer \\ 
        CBM & Condensed Baryonic Matter \\ 
        CERN & Conseil Européen de Recherche Nucléaire \\ 
        ChETEC & Chemical Elements as Tracers of the Evolution of the Cosmos \\ 
        CMS & particle physics experiment at LHC \\ 
        CP & charge conjugation and parity \\ 
        CRYRING@FAIR & Low-energy storage ring for heavy ions \\ 
        ECFA & European Committee for Future Accelerators \\ 
        ECT* & European Centre for Theoretical studies in nuclear physics and related areas \\ 
        EDM & electric dipole moment \\ 
        EFT & effective field theory \\ 
        EIC & electron ion collider \\ 
        ELENA & extra low energy antiproton ring \\ 
        ELI-NP & Extreme Light Infrastructure – Nuclear Physics \\ 
        ELSA & Electron Stretcher and Accelerator \\ 
        ESFRI & European Strategic Forum for Research Infrastructures \\ 
        ESR & experimental storage ring \\ 
        ESS & European Spallation Source \\ 
        FAIR & Facility for Antiproton and Ion Research \\ 
        FCC & Future Circular Collider \\ 
        FLAIR & Facility for Low-energy Antiproton and Ion Research \\ 
        GANIL & Grand Accélérateur National d’Ions Lourds \\ 
        GSI & Gesellschaft für Schwerionenforschung \\ 
        HADES & High Acceptance Di-Electron Spectrometer \\ 
        HIE-ISOLDE & high intensity and energy upgrade of ISOLDE \\ 
        IFMIF & International Fusion Materials Irradiation Facility \\ 
        IFMIF-DONES & IFMIF - DEMO-Oriented Neutron Source \\ 
        ILL & Institut von Laue – Langevin \\ 
               ISOL & isotope separation on-line \\ 
        ISOL@MYRRHA & isotope separation on-line at MYRRHA \\ 
        ISOLDE & ion separator on-line at CERN \\ 
        LHC & Large Hadron Collider \\ 
        LHCb & detector at LHC \\ 
        MAMI & Mainz Mikrotron \\ 
       MEDICIS & MEDical Isotopes Collected from ISOLDE \\ 
        MESA & Mainz Energy-Recovery Superconducting Accelerator \\ 
            MYRRHA & multi-purpose research reactor for high-tech applications \\ 
    \end{tabular}
\end{longtable}
        
 \clearpage\newpage       
\begin{longtable}[l]{ c  c }%
    \raggedleft
    \begin{tabular}{ll}
        n\_TOF & neutron time-of-flight facility at CERN\\ 
        PRISMAP & European medical radionuclides programme \\ 
        PSI & Paul Scherrer Institut \\ 
        QCD & quantum chromodynamics \\ 
        QED & quantum electrodynamics \\ 
        QGP & quark gluon plasma \\ 
        SDG & Sustainable Development Goal \\ 
        SM & standard model / shell model \\ 
        SPIRAL & Système de Production d'Ions Radioactifs Accélérés en Ligne \\ 
        SPS & Super Proton Synchrotron \\ 
        STRONG2020 & The strong interaction at the frontier of knowledge: fundamental research\\
        & and applications\\ 
    \end{tabular}
\end{longtable}
\end{document}